\documentclass[aps,letterpaper,amsmath,prb,twocolumn,superscriptaddress]{revtex4-1}
\usepackage[normalem]{ulem}
\usepackage{braket}
\usepackage[]{amsmath}
\usepackage[]{color}
\usepackage[]{graphicx}
\usepackage{bm}
\usepackage{subfigure}
\usepackage[percent]{overpic}
\usepackage{multirow}
\usepackage{bbold}

\graphicspath{{pics/}}

\begin{document}
\title{Cluster size convergence of the density matrix embedding theory and its dynamical cluster formulation: a study with an auxiliary-field quantum Monte Carlo solver}

\author{Bo-Xiao Zheng}
\affiliation{Department of Chemistry, Princeton University, New Jersey 08544, United States}
\affiliation{Division of Chemistry and Chemical Engineering, California Institute of Technology, Pasadena, California 91125, United States}
\author{Joshua S. Kretchmer}
\affiliation{Division of Chemistry and Chemical Engineering, California Institute of Technology, Pasadena, California 91125, United States}
\author{Hao Shi}
\affiliation{Department of Physics, The College of William and Mary, Williamsburg, Virginia 23187, United States}
\author{Shiwei Zhang}
\affiliation{Department of Physics, The College of William and Mary, Williamsburg, Virginia 23187, United States}
\author{Garnet Kin-Lic Chan}
\email{gkc1000@gmail.com}
\affiliation{Division of Chemistry and Chemical Engineering, California Institute of Technology, Pasadena, California 91125, United States}


\begin{abstract}
We investigate the cluster size convergence of the energy and observables using two forms
of density matrix embedding theory (DMET): the original cluster form (CDMET) and
a new formulation motivated by the dynamical cluster approximation (DCA-DMET).
Both methods are applied to the half-filled one- and two-dimensional Hubbard models
using a sign-problem free auxiliary-field quantum Monte Carlo (AFQMC) impurity solver, which
allows for the treatment of large impurity clusters of up to 100 sites. While 
CDMET is more accurate at smaller impurity cluster sizes, DCA-DMET exhibits faster asymptotic convergence towards the thermodynamic limit (TDL).
We use our two formulations to produce new accurate estimates
for the energy and local moment of the two-dimensional Hubbard model for  $U/t=2, 4, 6$. 
These results compare favourably with the best data available in literature, and help resolve earlier
uncertainties in the moment for $U/t=2$.
\end{abstract}
\maketitle


\section{Introduction}

Quantum embedding methods are a class of numerical techniques that help 
with simulating the physics of large and bulk interacting quantum systems. To reach the thermodynamic limit (TDL), one typically  
considers finite sized clusters of increasing sizes under some choice of boundary conditions, followed by a finite size scaling of the observables.
Embedding methods accelerate the finite size convergence, by mapping the
bulk problem onto an auxiliary impurity model, where a small cluster of the physical interacting sites are coupled to special ``bath sites'' that
mimic the effects of the neglected environment.

Dynamical mean-field theory (DMFT) and its cluster extensions~\cite{Georges1992,Georges1996,Maier2005a,Kotliar2006}, and the more
recent density matrix embedding theory (DMET) studied in this work~\cite{Knizia2012,Knizia2013,Wouters2016}, are two embedding methods of this kind. 
The bath sites in DMET~\cite{Knizia2012,Knizia2013} are constructed to capture entanglement between the bulk environment and the impurity cluster.
The entanglement-based construction ensures that the number of bath sites is at most equal to the number of impurity sites,
unlike the formally infinite bath representation that arises in DMFT methods.
Cluster DMET (CDMET) has been successfully applied to fermion and spin lattice models~\cite{Knizia2012,Booth2013,Chen2014,Bulik2014,Zheng2016},
as well as ab-initio molecular and condensed phase systems~\cite{Knizia2013,bulik2014electron,Tsuchimochi2015,Wouters2016}. 
In prior work~\cite{Zheng2016}, we showed that finite-size scaling of observables computed from quite small DMET impurity clusters 
can yield good estimates of the bulk observables. For example, 
in a study of the ground-state phase diagram of the 2D square-lattice Hubbard model, extrapolations from
clusters of only up to 16 sites already yielded a per-site energy accuracy
at half-filling of between $0.0003t$ ($U/t=2$) to $0.001t$ ($U/t=12$)~\cite{Zheng2016},
comparable with the best existing benchmark results~\cite{LeBlanc2015}. Nonetheless, the small sizes of these clusters leaves
open the possibility for a more detailed analysis of finite-size scaling in DMET. This is the question we revisit in
the present work, in the context of the half-filled 1D and 2D square lattice Hubbard models.

We have used exact diagonalization and density matrix renormalization group (DMRG) solvers
in earlier DMET work on Hubbard models, focusing on treating parts of the phase diagram where quantum Monte Carlo methods have a sign problem.
In the current study of cluster size convergence we focus on half-filling, where no sign problem exists. By using an
efficient auxiliary-field quantum Monte Carlo (AFQMC) implementation\cite{Sugiyama1986,Zhang2013}, we are able to study DMET clusters with up to 100 impurity sites.
Using this solver further facilitates direct comparisons to earlier bare (i.e. not embedded) AFQMC calculations in the literature that used
very large clusters (with up to 1058 sites) with periodic (PBC), anti-periodic (APBC), modified (MBC), and twisted boundary (TBC)
conditions~\cite{Sorella2015,Qin2016}. The comparison
provides a direct demonstration of the benefits of embedding, versus simply modifying the boundary conditions.

The finite-size scaling relation for extensive quantities assumed in earlier CDMET work was a simple surface-to-volume
law ($O(1/L)$ for extensive quantities, with $L$ being the linear dimension of the cluster). 
This is the same scaling used in cellular dynamical mean-field theory (CDMFT). The surface error arises
because the quantum impurity Hamiltonian in both CDMET and CDMFT describes
an impurity cluster with open boundary conditions, where the coupling between the impurity and the bath occurs only 
for sites along the boundary of the cluster~\cite{Fisher1972,Maier2002}. The open boundary nature
of the cluster further yields the well-known translational invariance breaking for impurity observables.
In contrast, the dynamical cluster approximation (DCA)\cite{PhysRevB.58.R7475,PhysRevB.61.12739,fotso2012dynamical}, a widely used alternative to CDMFT, restores translational
invariance for impurity observables by modifying the cluster Hamiltonian to use PBC. As a result, 
DCA calculations of extensive quantities converge as $O(1/L^2)$, faster than in CDMFT~\cite{Biroli2002,Aryanpour2005,Biroli2005}.
In this work, we introduce the DCA analog of DMET, which we term DCA-DMET, that uses a similarly modified cluster Hamiltonian. This restores
translational invariance and reproduces
the faster  $O(1/L^2)$ convergence in extensive quantities within the DMET setting.

Using both the existing CDMET and the new DCA-DMET formulations, together with large impurity cluster sizes, we compute new estimates
of the TDL energies and spin-moments of the 1D and 2D Hubbard model at half-filling for $U/t=4, 8$ and $U/t=2, 4, 6$, respectively.
For the energies, our results provide high accuracy benchmarks with small error bars.
Converging finite-size effects for spin-moment has well-known pitfalls, and 
existing data in the literature do not always agree~\cite{Varney2009,Wang2014,LeBlanc2015,Sorella2015,Qin2016}.
Where agreement is observed, our new 
estimates confirm the existing data with comparable or improved error bars. In the
case of $U/t=2$ where severe finite size effects are found, our data resolves between the earlier estimates in the literature.

\section{Methods} \label{sec:methods}
In this section, we provide a self-contained description of the computational methods in this work. We first introduce DMET, with a focus on the 
original CDMET formulation in Sec.~\ref{sec:dmet}, and then describe the DCA 
extension of DMET, DCA-DMET, in Sec.~\ref{sec:dca}. In Sec.~\ref{sec:finite}, we discuss the theoretical basis and motivation for the 
cluster-size scaling used in this work.
Finally in Sec.~\ref{sec:afqmc}, 
we briefly introduce AFQMC as the impurity solver, and discuss how to formulate
the DMET impurity Hamiltonian so as to preserve particle-hole symmetry (which removes the sign problem at half-filling in the Hubbard model).

\subsection{CDMET} \label{sec:dmet}
The original CDMET algorithm has been outlined in various recent 
works~\cite{Knizia2012,Knizia2013,Zheng2016,Wouters2016}, with
slightly different formulations used for  lattice model and ab-initio Hamiltonians.
In this section, we describe the algorithm used here that employs
the non-interacting bath formulation of CDMET~\cite{Knizia2012,Wouters2016}, 
as found in our previous work on lattice models~\cite{LeBlanc2015,Zheng2016}.
When required, we will assume we are working with the Hubbard
model, whose Hamiltonian is given by
\begin{equation}
\label{eq:hubbard}
H=-\sum_{\langle ij\rangle\sigma}ta_{i\sigma}^\dag a_{j\sigma}+\sum_iUn_{i\uparrow}n_{i\downarrow}
\end{equation}
where $a_{i\sigma}^\dag$ ($a_{i\sigma}$) creates (destroys) an particle of spin $\sigma$ at site $i$, 
$\langle ij\rangle$ denotes nearest neighbors, and $n_{i\sigma}=a_{i\sigma}^\dag a_{i\sigma}$. 


In CDMET, the exact ground-state wavefunction and expectation values
of the interacting Hamiltonian, $H$, defined on the full lattice, are approximated by self-consistently
solving for the ground-state of two coupled model problems: ({i}) an interacting problem defined for a
quantum impurity, and ({ii}) an auxiliary non-interacting system defined on the original lattice. 
The quantum impurity model, with Hamiltonian $H_{\mathrm{imp}}$ and ground-state
$|\Psi\rangle$, consists of $N_{\mathrm{imp}}$ cluster sites coupled to
$N_{\mathrm{imp}}$ bath sites. The bath sites are obtained from the Schmidt
decomposition\cite{ekert1995entangled} of the ground-state, $|\Phi\rangle$,
of the auxiliary non-interacting system, with Hamiltonian $h$. A self-consistency
condition on the one-particle reduced density matrix then links the two model
problems.

To define the Hamiltonian $h$, we first partition the total lattice into $N_c = N/N_{\mathrm{imp}}$ fragments, termed impurity clusters, which tile the full lattice.
We then choose the auxiliary Hamiltonian $h$ to be a quadratic Hamiltonian of the form
\begin{equation}
h=h_0+u
\label{eq:h_mf}
\end{equation}
where $h_0$ is the one-body part of $H$ (the hopping term of the Hubbard Hamiltonian in Eq. \eqref{eq:hubbard}) and $u$ is the local \textit{correlation potential}.
In this work, we do not consider superconducting phases and we choose to preserve $S_z$ symmetry. This restricts $u$ to be number conserving
and of the form
\begin{equation}
  u=\sum_C\sum_{i,j\in C}\sum_{\sigma}u_{ij\sigma}a_{i\sigma}^{\dagger}a_{j\sigma}
  \label{eq:corr_pot}
\end{equation}
where $C$ indexes the $N_c$ clusters and $\sum_{i,j\in C}$ is restricted to the sites of cluster $C$.
 The correlation potential approximates the effect of the
local Coulomb interaction within each cluster for the auxiliary problem and is a kind of ``mean-field''. The elements  $u_{ij\sigma}$ are
determined through the self-consistency condition described below. As we vary $u_{ij\uparrow}$ and $u_{ij\downarrow}$ 
independently, this allows for $S^2$ symmetry breaking.

The bath states that define the quantum impurity model associated with cluster $C$ are obtained from the ground-state of $h$, 
$|{\Phi}\rangle$, which takes the form of a simple Slater determinant. The bath states can be constructed from $|\Phi\rangle$ in 
several mathematically equivalent ways. Here, we use a singular value decomposition of
(part of) the one-particle density matrix $\rho_\Phi^\sigma$, computed from $|\Phi\rangle$, with elements $[{\rho_\Phi^{\sigma}}]_{ij}=\langle\Phi|a_{i\sigma}^\dag a_{j\sigma}|\Phi\rangle$
defined over the entire lattice. 
For a given impurity cluster $C$, $\rho^\sigma$ can be partitioned 
into a $N_{\mathrm{imp}}\times N_{\mathrm{imp}}$
impurity block, a $(N-N_{\mathrm{imp}})\times(N-N_{\mathrm{imp}})$ environment block, and $N_{\mathrm{imp}}\times(N-N_{\mathrm{imp}})$ off-diagonal coupling blocks,
\begin{equation}
\rho_\Phi^{\sigma}\equiv
\begin{bmatrix}
    \rho_{\text{imp}}^\sigma&\rho_{\text{c}}^\sigma\\
    \rho_{\text{c}}^{\sigma\dagger} & \rho_{\text{env}}^\sigma
    \end{bmatrix}.
\end{equation}
The bath spin-orbitals associated with impurity cluster $C$ and spin $\sigma$ are obtained by performing a singular value decomposition of the coupling block
\begin{equation}
  \rho_{\text{c}}^\sigma=R_{\text{imp}}^\sigma\Sigma^\sigma R_{\text{bath}}^{\sigma\dagger}
  \label{eq:bath}
\end{equation}
where $R_{\text{bath}}$ is the $(N-N_{\mathrm{imp}})\times N_{\mathrm{imp}}$ coefficient matrix defining the $N_{\mathrm{imp}}$ single-particle bath
spin-orbitals as a linear combination of the environment lattice sites. The impurity model derived from cluster $C$ thus consists of the  $2N_{\mathrm{imp}}$ spin-orbitals associated with the original
sites restricted to the impurity cluster, and the $2N_{\mathrm{imp}}$ delocalized, environmental bath spin-orbitals (where the factor of two accounts for both up and down spins).
In principle, we would need to construct an impurity model for each cluster $C$, but because of translational symmetry in the
Hubbard model, all clusters are equivalent, thus only one cluster, say $C=0$, is used as the impurity.

In the non-interacting bath CDMET formulation, the Hamiltonian of the impurity problem, $H_{\mathrm{imp}}$, is obtained by projecting
an Anderson-like Hamiltonian, $H_{\mathrm{NI}}$ (where $\mathrm{NI}$ denotes the non-interacting formulation), defined on the full lattice, into the Fock space spanned by the impurity and bath states.
The Hamiltonian $H_{\mathrm{NI}}$ differs from the original Hubbard Hamiltonian
in that the  interaction terms in the environment are replaced with the one-body
correlation potential, such that
\begin{eqnarray}
  H_{\mathrm{NI}}&=&h_0+U\sum_{i\in C=0}n_{i\uparrow}n_{i\downarrow}+\sum_{C\neq0}\sum_{i,j\in C}\sum_{\sigma}u_{ij\sigma}a_{i\sigma}^\dag a_{j\sigma}\nonumber\\
&\equiv&h_0+V_{\mathrm{imp}}+u_{\mathrm{env}}
\end{eqnarray}
where $C=0$ corresponds to the impurity cluster and the set $\{C\neq0\}$ corresponds to the clusters that comprise the environment.
Due to the simple structure of the Schmidt decomposition of $|\Phi\rangle$, the projection of $H_{\mathrm{NI}}$ into the impurity plus bath Fock space
can equivalently be performed by a rotation of the one-particle basis~\cite{Knizia2012,Knizia2013,Wouters2016}, giving
%
\begin{equation}
H_{\mathrm{imp}}=\bar{h}+V_{\mathrm{imp}}
\label{eq:h_imp}
\end{equation}
where
\begin{equation}
\bar{h}=\sum_{pq}\sum_{\sigma}\bar{h}_{pq\sigma}a_{{p\sigma}}^\dag a_{{q\sigma}}.
\label{eqn:hbar0}
\end{equation}
The indices $p\sigma$ and $q\sigma$  label the impurity and bath spin-orbitals, 
and the matrix $\bar{h}_\sigma$ is defined as
\begin{equation}
\bar{h}_{\sigma}=R^{\sigma\dag}\left(h_0+u_{\mathrm{env}}^{\sigma}\right)R^{\sigma}
\label{eqn:hbarsigma}
\end{equation}
where
\begin{equation}
R^{\sigma}=
\begin{bmatrix}
    \mathbb{1}_{N_{\mathrm{imp}}\times N_{\mathrm{imp}}}&0\\
    0 & R^{\sigma}_{\mathrm{bath}}
    \end{bmatrix}
\end{equation}
is the rotation matrix from the original lattice site basis to the basis of single-particle impurity and bath states.
It is important to note that the impurity states are the same in either basis as denoted by the identity in the upper-left
block of $R^\sigma$.

To compute the ground-state of the impurity model Hamiltonian ${H}_\text{imp}$, we can choose from a wide range of ground state solvers
depending on the nature of the problem as well as the cost and accuracy requirements. Previous DMET calculations
have used exact diagonalization and DMRG impurity solvers for strongly correlated problems
~\cite{Knizia2012,Chen2014,Zheng2016,Bulik2014}, and coupled cluster theory for more weakly correlated,
ab-initio calculations~\cite{bulik2014electron,Wouters2016}.
In this work, we use an auxiliary-field quantum Monte Carlo (AFQMC)~\cite{Sugiyama1986,Zhang2013,Shi2015} solver, which does not have a sign problem
at half-filling in the Hubbard model that we study here. This solver is discussed in more detail in Section~\ref{sec:afqmc}.

As described above,  the elements of the correlation potential $u$ are determined by a self-consistent procedure.
We maximize the ``similarity'' between the lattice uncorrelated wavefunction $|\Phi\rangle$ and the impurity model correlated wavefunction $|\Psi\rangle$, 
measured by the Frobenius norm of the difference between their one-body density matrices, projected to the impurity model (this is the
``fragment plus bath'' cost function in Ref.~\cite{Wouters2016})
\begin{equation}
  \min_u f(u)=\sqrt{\sum_{ij\sigma}\{[R^{\sigma\dagger}\rho_{\Phi}^\sigma(u)R^{\sigma}]_{ij}-[\rho_{\Psi}^\sigma(u)]_{ij}\}^2}
  \label{eq:cost}
\end{equation}
where the elements $[\rho_{\Psi}^\sigma]_{pq} = \langle \Psi|a_{p\sigma}^\dag a_{q\sigma}|\Psi\rangle$.
Because direct optimization of the functional $f(u)$ requires computing the gradient of the correlated wavefunction $d\Psi/du$, a self-consistent
iteration is used: when optimizing $f(u)$, $|\Psi\rangle$ is fixed; the optimal $u$ is then used to update $|\Phi\rangle$, the impurity Hamiltonian $H_{\text{imp}}$,
and thus $|\Psi\rangle$.

In a summary, the DMET calculations in this work proceed via the following steps:
\begin{enumerate}
  \item we choose an initial guess for the correlation potential $u$;
  \item we solve for the lattice Hamiltonian $h$ (Eq.~(\ref{eq:h_mf})) to obtain the lattice wavefunction $|\Phi\rangle$;\label{item:mf}
  \item we construct the impurity model Hamiltonian using Eq.~(\ref{eq:h_imp});
  \item we use the AFQMC impurity solver to compute the ground state of the impurity model, $|\Psi\rangle$, and construct the one-body density matrix $\rho_{\Psi}$;
  \item we minimize $f(u)$ in Eq.~(\ref{eq:cost}), with $\rho_{\Psi}$ fixed, to obtain the new correlation potential $u^\prime$;
  \item if $||u-u^\prime||_\infty > \varepsilon_0$, the convergence threshold, we set $u=u^\prime$ and go to step~\ref{item:mf}; otherwise the DMET calculation is converged.
    Here the infinite norm $||\cdot||_\infty$ simply takes the maximum absolute value of a matrix.
\end{enumerate}

We now briefly discuss how to compute the energy and other observables in DMET.
The energy per impurity cluster, $E / N_c$, where $E$ is the total energy
of the lattice and $N_c$ is the number of impurity clusters,
can be defined as the sum of the impurity internal energy and the coupling energy with the environment~\cite{Knizia2013,Wouters2016}. 
Due to the local nature of the interactions in the Hubbard model, one arrives at the simplified expression,
\begin{equation}
  \begin{split}
    e=\frac{E}{N_c}&=\sum_{p\in\text{imp},q,\sigma}\bar{h}_{0,pq\sigma}\rho^{\Psi}_{pq\sigma}+\sum_{p\in\text{imp}}U\langle n_{p\uparrow}n_{p\downarrow}\rangle_\Psi\\
    &=\sum_{p\in\text{imp},q,\sigma}\bar{h}_{0,pq\sigma}\rho^{\Psi}_{pq\sigma}+(E_{\text{imp}}-\sum_{p,q,\sigma}\bar{h}_{pq\sigma}\rho^{\Psi}_{pq\sigma})\\
    &=E_{\text{imp}}-\sum_{p\in\text{bath},q}\bar{h}_{pq\sigma}\rho^{\Psi}_{pq\sigma}
  \end{split}
  \label{eq:energy}
\end{equation}
where $p,q$ range only over the impurity and bath orbitals, 
$\bar{h}_{0,\sigma}=R^{\sigma\dagger}h_0R^{\sigma}$ is the bare
one-particle Hamiltonian  projected to the impurity model, 
and
$E_{\text{imp}}=\langle\Psi|{H}_{\text{imp}}|\Psi\rangle$ is the ground-state energy of the impurity model. 
Note that Eq.~(\ref{eq:energy}) only explicitly involves the {\it one-particle} density matrix of the impurity model. This
is a significant benefit as it reduces the computational cost in the AFQMC solver.

\begin{figure}[htpb]
  \centering
  \subfigure[]{
    \includegraphics[width=\columnwidth]{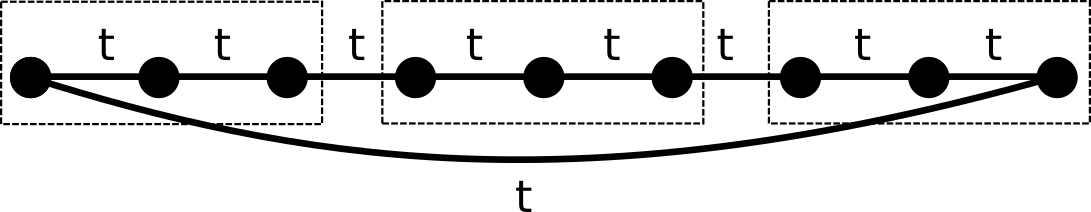}
    \label{fig:lattice}
  }
  \subfigure[]{
    \includegraphics[width=0.4\columnwidth]{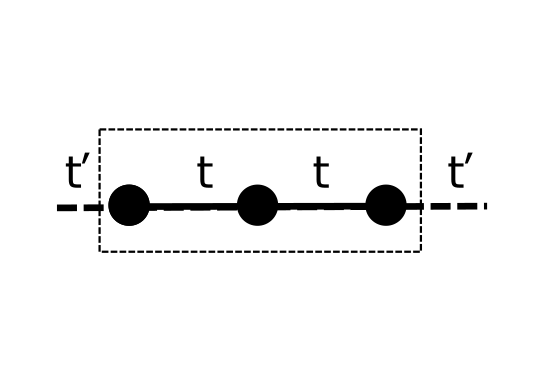}
    \label{fig:cdmet}  
  }
  \subfigure[]{
    \includegraphics[width=0.4\columnwidth]{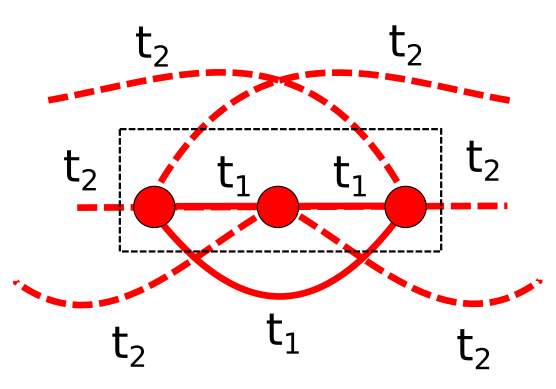}
    \label{fig:dca}  
  }

  \caption{Translational symmetry in DMET. (a) The original lattice with translational symmetry, divided into 3 supercells. 
    (b) The CDMET impurity cluster with broken intracluster translational symmetry, between the center site and the edge sites.
    (c) The DCA-DMET impurity cluster restores the intracluster translational symmetry through a basis transformation and interaction coarse-graining.}
  \label{fig:translational}
\end{figure}

Local observables, such as charge and spin densities as well as correlation functions, can be extracted directly from the
correlated impurity wavefunction $|\Psi\rangle$. These quantities, however, are most accurate when measured within the impurity cluster,
where interactions are properly treated.  While CDMET preserves translational symmetry between supercells,
the intracluster translational symmetry is generally broken, as illustrated in Fig.~\ref{fig:translational}. This leads to some ambiguity in defining the local order
parameters. We illustrate the magnitude of this symmetry breaking and the consequences of
different definitions in Sec.~\ref{sec:results}; in Sec.~\ref{sec:dca}, we introduce the DCA-DMET
formulation which restores translational symmetry. 

\subsection{DCA-DMET} \label{sec:dca}

\begin{figure}[htpb]
  \centering
  \includegraphics[width=\columnwidth]{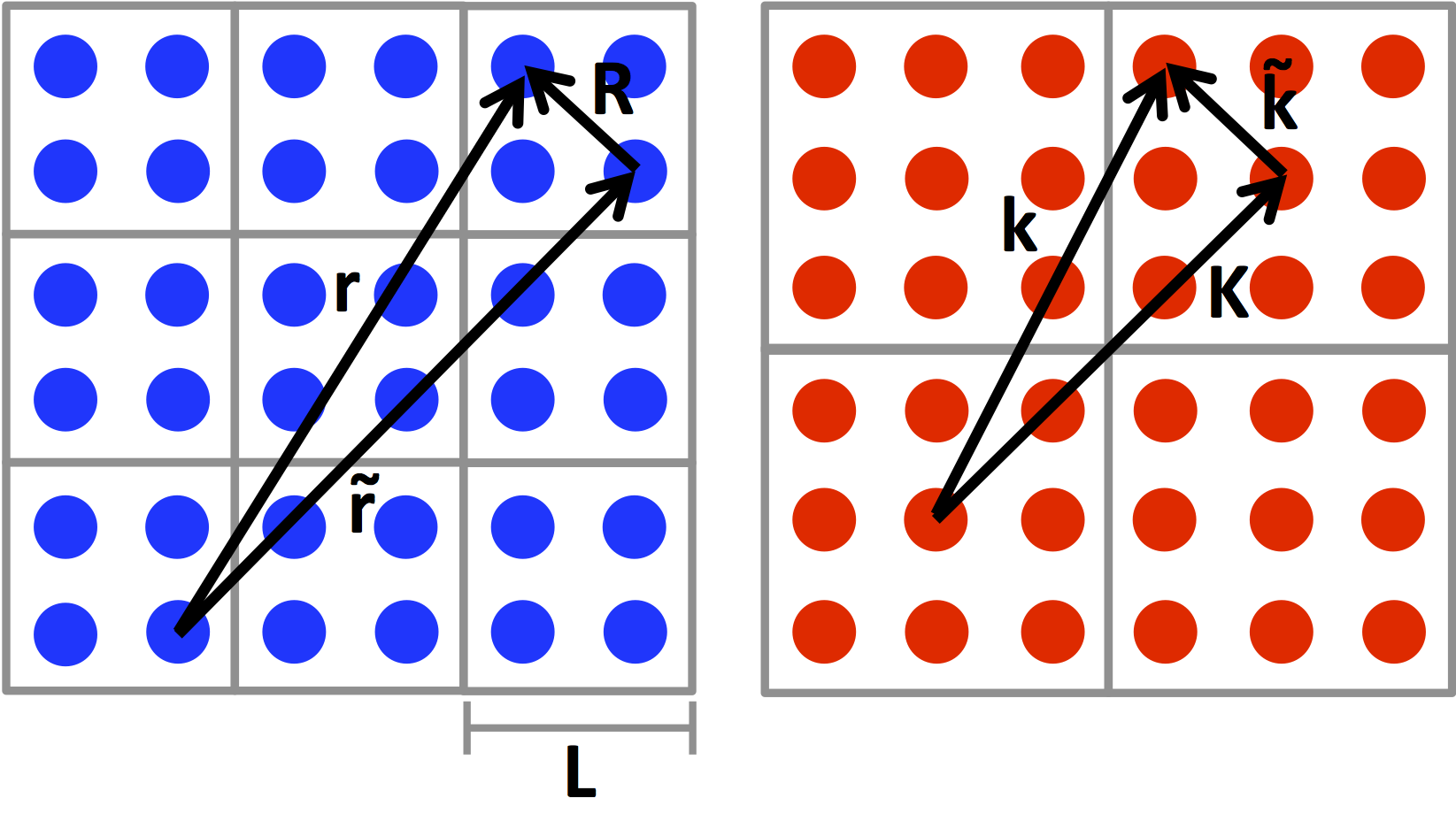}
  \caption{Definition of the real (left) and reciprocal (right) lattice vectors for the DCA transformation for a ``hypercubic" cluster with $L=2$. The inter-cluster component of the real lattice vector, $\tilde{\mathbf{r}}$, labels the origin of the cluster, and the intra-cluster component, $\mathbf{R}$, labels the site within the cluster. The reciprocal space of $\tilde{\mathbf{r}}$ and $\mathbf{R}$ are labeled by $\tilde{\mathbf{k}}$ and $\mathbf{K}$, respectively.}
  \label{fig:dca_vec}
\end{figure}

In CDMET, the form of the Hamiltonian within the impurity sites is simply the original lattice Hamiltonian restricted
to the impurity sites.
In DCA-DMET, we transform the lattice Hamiltonian such that the restriction to a finite cluster retains a periodic boundary
within the cluster, thus restoring the intracluster
translational symmetry (Fig.~\ref{fig:translational}).
The DCA transformation involves two  steps: a basis rotation which redefines the lattice single-particle Hamiltonian, and  a coarse graining of
the two-particle interaction~\cite{PhysRevB.58.R7475,PhysRevB.61.12739,Maier2005a,Potthoff2007}.

To introduce the DCA transformation, we first define  the intra- and inter-cluster components of the real and reciprocal lattice vectors (Fig.~\ref{fig:dca_vec}),
\begin{equation}
  \mathbf{r} = \mathbf{R}+\tilde{\mathbf{r}}, \ \ \mathbf{k} = \mathbf{K}+ \tilde{\mathbf{k}}.
  \label{eq:DCAdecomp}
\end{equation}
For simplicity we will assume ``hypercubic'' lattices (in arbitrary dimension) with orthogonal unit lattice vectors with linear dimension $L$, and
``hypercubic'' clusters with linear dimension $L_c$. The corresponding super-cell lattice  then has orthogonal lattice vectors of magnitude $L_c$, and the
total number of supercells along each linear dimension is $L/L_c$.
The intracluster lattice vector,  $\mathbf{R} = (R_1, R_2, \ldots)$
and reciprocal lattice vector $\mathbf{K} = 2\pi/L_c ( N_1,  N_2, \ldots)$ where $0\le  R_i, N_i < L_c; \ R_i, N_i \in \mathbb{Z}$,
and intercluster components $\tilde{\mathbf{r}}=L_c (\tilde{r}_1, \tilde{r}_2 \ldots)$, $\tilde{\mathbf{k}}=2\pi/L ( \tilde{n}_1,  \tilde{n}_2, \ldots)$,
with $0 \le \tilde{r}_i,\tilde{n}_i < L/L_c; \ \tilde{r}, \tilde{n} \in \mathbb{Z}$, are uniquely defined for any $\mathbf{r}$ and $\mathbf{k}$.

Our goal is to obtain a Hamiltonian which is {\it jointly} periodic in the intracluster and intercluster lattice vectors, $\mathbf{R}$ and $\tilde{\mathbf{r}}$.
Such a jointly periodic  basis  is provided by the product functions $e^{-i \tilde{\mathbf{k}} \cdot \tilde{\mathbf{r}}} e^{-i \mathbf{K} \cdot{\mathbf{R}}}$. 
From $h$ defined in reciprocal space, $h=\sum_\mathbf{k} h (\mathbf{k}) a_\mathbf{k}^{\dagger}a_\mathbf{k}$,
  and with the mapping in Eq.~(\ref{eq:DCAdecomp}),
  we identify the diagonal DCA Hamiltonian matrix elements in the jointly periodic basis as
  \begin{equation}
    h (\mathbf{k}) \to  h_{\mathrm{DCA}} (\mathbf{\tilde{k}, \mathbf{K}}).
  \end{equation}
  The inverse Fourier transformation then gives the DCA matrix elements on the real-space lattice. The Fourier transforms between
 the
   different single particle Hamiltonians  are summarized as:
\begin{equation}
  \begin{split}
  h(\mathbf{r})&\xrightarrow[]{e^{-i\mathbf{k}\cdot \mathbf{r}}}h(\mathbf{k})\xrightarrow[]{\mathbf{k}=\tilde{\mathbf{k}}+\mathbf{K}} h_\text{DCA}(\tilde{\mathbf{k}},\mathbf{K})\\
  &  \xrightarrow[]{e^{i\tilde{\mathbf{k}}\tilde{\mathbf{r}}}}
  \xrightarrow[]{e^{i\mathbf{K}\cdot\mathbf{R}}}h_{\text{DCA}}(\tilde{\mathbf{r}},\mathbf{R})
  \end{split}.
  \label{eq:DCAflow}
\end{equation}
The resultant real-space matrix elements, $h_{\text{DCA}}(\tilde{\mathbf{r}},\mathbf{R})$, thus only depend on the inter- and intra-cluster
separation between sites.

The transformation from $h(\mathbf{r}) \to h_\text{DCA}(\tilde{\mathbf{r}}, \mathbf{R})$ is simply a basis transformation of $h$, with the rotation matrix defined as~\cite{Potthoff2007} 
\begin{equation}
  U_{\mathbf{R}+\tilde{\mathbf{r}}, \mathbf{R}^\prime+\tilde{\mathbf{r}}^\prime}=\sum_{\mathbf{K},\tilde{\mathbf{k}}}e^{-i[\mathbf{K} \cdot (\mathbf{R}^\prime-\mathbf{R})+\tilde{\mathbf{k}}(\tilde{\mathbf{r}}^\prime-\tilde{\mathbf{r}})+\tilde{\mathbf{k}} \cdot \mathbf{R}^\prime]}.
  \label{eq:unitary}
\end{equation}
Viewing the DCA transformation as a basis rotation suggests that the same transformation should be extended to the interaction terms as well,
generating non-local interactions. 
However, in DCA one argues to use the ``coarse-grained'' interaction in momentum space, introducing a discrepancy at finite sizes which vanish as cluster size grows.
In the Hubbard model, the coarse-graining leaves
the local $Un_{i\uparrow}n_{i\downarrow}$ term unchanged in the transformed Hamiltonian. Note that the coarse-grained interaction is non-local if
transformed back to the original site basis using the rotation in Eq.~(\ref{eq:unitary}).

\subsection{Finite-size convergence}
\label{sec:finite}

We now analyze the cluster finite-size convergence of observables in CDMET and DCA-DMET in $d$ dimensions. For
the energy, we use a perturbation argument to obtain the leading term of the finite size scaling; for the more
complicated case of intensive observables, we suggest a plausible scaling form.

\begin{figure}[htpb]
  \centering
  \includegraphics[width=\columnwidth]{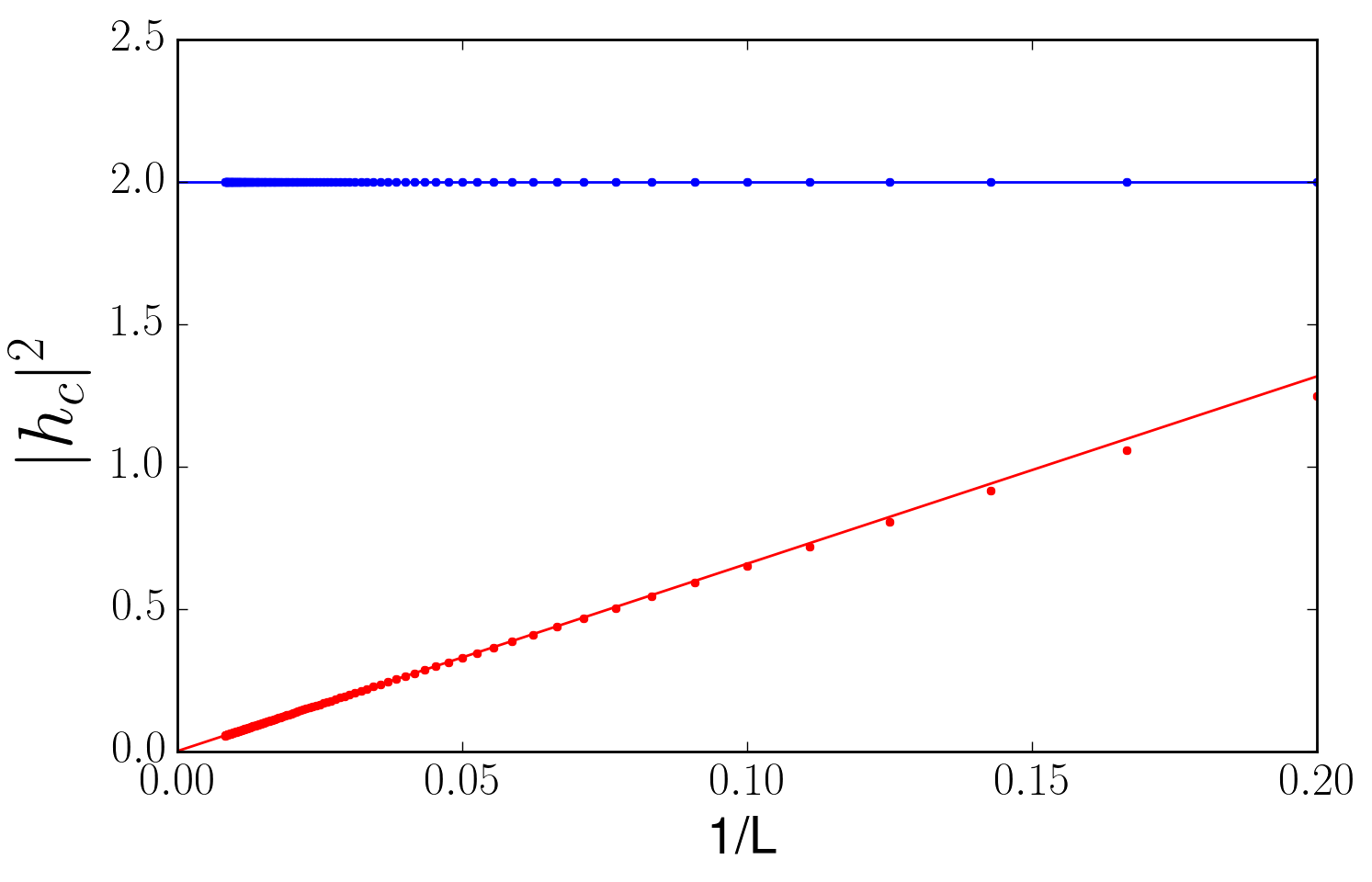}
  \caption{Sum-of-square of the one-body impurity-environment coupling Hamiltonian $|h_c|^2=\sum_{i\in C=0,j\in C^\prime\neq0}|h_{ij}|^2$
    for the CDMET and DCA formulations, in one-dimension. The fittings follow constant (CDMET) and 1/L (DCA) scalings, respectively.}
  \label{fig:coupling}
\end{figure}

We consider the following factors to derive the DMET finite-size scaling:
  (a) the open boundary  in CDMET; (b) the gapless spin excitations of quantum antiferromagnets;
  (c) the coupling between the impurity and bath; (d) the modification of the hoppings of the Hubbard Hamiltonian in DCA-DMET.

We start with the CDMET energy. We first consider the {\it bare} impurity cluster in CDMET
(i.e. without the bath) which is just the finite size truncation of the TDL system. For a gapped system, we expect an open boundary
to lead to a finite-size energy error (per site) proportional to the surface area to volume ratio~\cite{Fisher1972}, i.e. 
\begin{align}
  e(L) &= {e(\infty)} + \frac{a_0}{L} + \ldots \label{eq:open_FS}
  \end{align}
where $e(L)$ is the energy per site for an $L^d$ site cluster and $e(\infty)$ is the energy per site in the TDL.
If, in the TDL, there are gapless modes, a more careful analysis is required. The Hubbard model studied
here has gapless spin excitations. These yield a finite size error of $O(1/L^{d+1})$
in a cluster with PBC~\cite{Fisher1989,Hasenfratz1993,Huse1988,Sandvik1997}.
This is subleading to the surface finite size error introduced by the open boundary in Eq.~(\ref{eq:open_FS}) for $d>0$.

We next incorporate the CDMET bath coupling. Each
site on the impurity cluster boundary couples to the bath, yielding a total Hamiltonian coupling of $O(1)$
per boundary site (see Fig.~\ref{fig:coupling}). The
total ``perturbation'' to the bare impurity cluster Hamiltonian is then $O(L^{d-1})$, which leads to a first order energy correction per site
of
\begin{align}
{e(L)_\text{CDMET}} &=  {e(\infty)} + \frac{a_0'}{L}  + \ldots 
\end{align}
For the perfect DMET bath (derived from the exact auxiliary wavefunction), $a_0' = 0$, thus we
expect $a_0'$ to be small in practice. 

For DCA-DMET, the above argument must be modified in two ways: first, the impurity
cluster uses PBC, and second, the formulation modifies inter-cluster and 
intra-cluster hoppings. Similarly, we start with
the bare periodic impurity cluster (without any modification of the intra-cluster hoppings).
In the TDL, for a gapped state with short-range interactions, all correlation functions decay exponentially (e.g. Wannier functions are exponentially localized)
and we expect an exponential convergence of the energy with respect to cluster size. 
However, in the Hubbard model, the gapless spin excitations
give a finite-size energy error (per site) of $O(1/L^{d+1})$.
The leading order finite-size scaling for the bare periodic cluster
is thus expected to be
 \begin{align}
   e(L) &= e(\infty) + \frac{a_0}{L^{d+1}} + \ldots
   \end{align}

The DCA-DMET  Hamiltonian  modifies the periodic cluster 
Hamiltonian by changing both the intracluster and intercluster hopping terms.
The intracluster hopping terms are modified by a term of order
$O(1/L^2)$, and the intercluster hopping terms are modified so as to generate
a coupling between each site in the cluster and the bath with a total interaction strength of $O(1/L^2)$ (see Fig.~\ref{fig:coupling}).
Since there are $L^d$ sites
in the cluster,  the total
magnitude of the DCA-DMET perturbation (including the contributions of both intracluster and intercluster terms)
is $O(L^{d-2})$. 
For dimension 1, the perturbation and impurity-bath coupling give the leading term in the finite-size error, while in dimension 2, they give
a contribution with the same scaling as the contribution of the gapless modes. Thus combining the three sources of finite-size error we expect in 1 and 2 dimensions
a scaling of the form,
\begin{align}
e(L)_\text{DCA-DMET} &= {e(\infty)} + \frac{a_0'}{L^2} + \ldots
  \end{align}
Note that the scaling of the CDMET and DCA-DMET energies is the same  as is found for CDMFT and DCA.

The finite size scaling of intensive quantities is more tricky to analyze~\cite{Maier2002}. 
For an observable $Q$ we have the relation
$\langle Q \rangle = \lim_{r\to\infty} \langle Q(0) Q(r) \rangle^{1/2} $, where $\langle Q(0) Q(r)\rangle$ is a correlation function.
It is often argued that the error in $\langle Q \rangle$ in 
a {large} finite cluster  behaves like 
\begin{align}
  \Delta Q  \sim  [\langle Q(0) Q(R)\rangle^{1/2} - \langle Q(0) Q(\infty)\rangle^{1/2}]
  \label{eq:dQ}
\end{align}
where $R$ is the largest length in the cluster~\cite{Huse1988} $\sim L/2$.
{For CDMET, where the cluster is only coupled to the symmetry-broken
  bath at the boundary, we assume the form in Eq.(~\ref{eq:dQ}) holds, with additional corrections from the
system size, expanded as a Taylor series}
\begin{align}
  \Delta Q =  \left(a + \frac{b}{L} + \ldots\right) [\langle Q(0) Q(R)\rangle^{1/2} - \langle Q(0) Q(\infty)\rangle^{1/2}] \label{eq:intensive}
\end{align}
Eq.~(\ref{eq:intensive}) is a heuristic form and its correctness will be assessed in our numerical results.
For the local magnetic moment $m = \langle S_z\rangle$, the correlation function  $\langle S_z(0) S_z(r)\rangle$ behaves at large $r$ like $a \sqrt{\ln r} /r$ in the  1D Hubbard model and $a+b/r$ in the 2D square-lattice Hubbard model at half-filling. 
Consequently, we assume a scaling form in 1D of 
 \begin{align}
   m(L)_\text{CDMET} = \sqrt{\frac{\sqrt{\ln L/2}}{L/2}} \left(a + \frac{b}{L} + \ldots\right) \label{eq:1dmscaling}
 \end{align}
and in 2D of
\begin{align}
  m(L)_\text{CDMET} = a + \frac{b}{L} + \frac{c}{L^2} + \ldots. \label{eq:2dmscaling} 
\end{align}

For DCA-DMET, however, every impurity site, not just those at the boundary, are coupled to a set of bath orbitals, which provide  a
symmetry-breaking field. This means that there is no simple connection to the correlation function of the system.
Therefore, we use an empirical form for the DCA-DMET magnetic moment in both one- and two-dimensions,
\begin{align}
  m(L)_\text{DCA-DMET} = a + \frac{b}{L} + \frac{c}{L^2} + \ldots.
  \label{eq:dcamscaling}
\end{align}


\subsection{AFQMC} \label{sec:afqmc}

In this work, we use AFQMC~\cite{Sugiyama1986,Zhang2013,Shi2015,Shi2016} to solve for the ground state of the impurity model. We briefly introduce the general ideas here,
while details of the algorithm can be found in Ref.~\cite{Zhang2013,Shi2015,Shi2016}.
AFQMC obtains the ground state of a fermionic Hamiltonian through the imaginary time evolution of a trial wavefunction
\begin{equation}
  |\Psi_0\rangle \propto \lim_{\beta\rightarrow\infty}e^{-\beta H}|\Psi_T\rangle
  \label{eq:imag_time}
\end{equation}
The time evolution is carried out using the second-order Trotter-Suzuki decomposition, 
\begin{equation}
  e^{-\beta H}=(e^{-\tau H})^n=(e^{-\frac{\tau}{2} H_1}e^{-\tau H_2}e^{-\frac{\tau}{2} H_1})^n+O(\beta\tau^2)
\end{equation}
where $H_1$ and $H_2$ are the one- and two-body parts of the Hamiltonian.

Given any Slater determinant $|\Psi\rangle=|\phi_{1\uparrow}\dots\phi_{N\uparrow}\rangle\otimes|\phi_{1\downarrow}\dots\phi_{N\downarrow}\rangle$ and any one-body operator
\begin{equation}
  K=\sum_{ij\sigma}k_{ij\sigma}a_{i\sigma}^{\dagger}a_{j\sigma}
\end{equation}
the canonical transformation $e^{K}|\Psi\rangle$ can be carried out exactly, giving another Slater determinant
$|\Psi^\prime\rangle =e^{K}|\Psi\rangle=
|\phi_{1\uparrow}^\prime\dots\phi_{N\uparrow}^\prime\rangle\otimes|\phi_{1\downarrow}^\prime\dots\phi_{N\downarrow}^\prime\rangle$ with 
the coefficient matrix
\begin{equation}
  \Phi_\sigma^\prime=\left(
  \phi_{1\sigma}^\prime,\dots,\phi_{N\sigma}^\prime
  \right)=e^{k_\sigma}\Phi_\sigma
  \label{eq:boost}
\end{equation}
The matrix multiplication in Eq.~(\ref{eq:boost}) gives the $O(N^3)$ scaling of the AFQMC algorithm (where $N$ is system size).
Starting with a Slater determinant as the trial wavefunction $|\Psi_T\rangle$, the propagation of the one-body Hamiltonian can be treated using
Eq.~(\ref{eq:boost}), by letting $K=-\frac{\tau}{2}H_1$.


The propagation of the two-body part of the Hamiltonian is rewritten
as a sum over one-body propagations using a Hubbard-Stratonovich
transformation. For the Hubbard model, we use the discrete form of this transformation,
\begin{align}
  e^{-\tau Un_{i\uparrow}n_{i\downarrow}}&=e^{-\tau U(n_{i\uparrow}+n_{i\downarrow})/2}\sum_{x_i=\pm1}\frac{1}{2}e^{\gamma x_i(n_{i\uparrow}-n_{i\downarrow})}\notag\\
  &=\sum_{x_i=\pm1} e^{ V(x_i,\tau)}
  \label{eq:sdecomp}
\end{align}
where $x_i$ is a binary auxiliary field, and $\cosh{\gamma}=\exp(-\tau U/2)$. Eq.~(\ref{eq:sdecomp}) is often termed ``spin decomposition'', in contrast to another
possible formed called ``charge decomposition''. The choice of different transformations does affect the accuracy and efficiency in AFQMC calculations~\cite{Shi2013}.

The auxiliary field $x_i$ 
is sampled to obtain a stochastic representation of the propagation, 
and thus of the ground state wavefunction $|\Psi_0\rangle$ as a sum of walkers.
General observables are calculated from the pure estimator, where the summations are similarly sampled,
\begin{widetext}
  \begin{equation}
    \langle \hat{O}\rangle =\lim_{n\rightarrow \infty}\frac{\sum_{{x}_1} \dots \sum_{{x}_n} \sum_{{x}_1^\prime}\dots \sum_{{x}_n^\prime}
    \langle \Psi_T|\prod_{j=1}^n(e^{-\frac{\tau}{2} H_1}e^{-\hat{V}({x}_j^\prime,\tau)}e^{-\frac{\tau}{2} H_1})\hat{O}
    \prod_{i=1}^n(e^{-\frac{\tau}{2} H_1}e^{-\hat{V}({x}_i,\tau)}e^{-\frac{\tau}{2} H_1})|\Psi_T\rangle }
    {\sum_{{x}_1} \dots \sum_{{x}_n} \sum_{{x}_1^\prime}\dots \sum_{{x}_n^\prime}
    \langle \Psi_T|\prod_{j=1}^n(e^{-\frac{\tau}{2} H_1}e^{-\hat{V}({x}_j^\prime,\tau)}e^{-\frac{\tau}{2} H_1})
    \prod_{i=1}^n(e^{-\frac{\tau}{2} H_1}e^{-\hat{V}({x}_i,\tau)}e^{-\frac{\tau}{2} H_1})|\Psi_T\rangle }
  \label{eq:AFQMCobs}
\end{equation}
\end{widetext}
The energy may be computed using a simpler estimator (the mixed estimator) where the propagation of the bra is omitted.




The sign problem arises because the individual terms in the denominator in Eq.~(\ref{eq:AFQMCobs}) can be both positive and negative and
lead to a vanishing average with infinite variance. When there is a sign problem,
a constrained path approximation can be invoked in the calculation which removes the problem with a gauge condition using a trial wave function
~\cite{PhysRevLett.90.136401,PhysRevB.55.7464,Zhang1995}.
In certain models, however, such as the half-filled repulsive Hubbard model on a bipartite lattice, the sign-problem 
does not arise because the overlap between every walker and the trial wavefunction is guaranteed to be non-negative.
It turns out that, in these models, the DMET impurity Hamiltonian
is also sign-problem free as long as  certain constraints are enforced on the correlation potential.
For the half-filled Hubbard model on a bipartite lattice, the condition is
\begin{equation}
  u_{ij,\uparrow}+(-)^{i+j}u_{ij,\downarrow}=\delta_{ij}U
  \label{eq:phsymm}
\end{equation}
The parity term $(-)^{i+j}$ takes opposite signs for the two sublattices. The derivation of this constraint
is given in Appendix~\ref{sec:phsymm}.

In this work, we use the AFQMC implementation described in Ref.\cite{Zhang2013,Shi2015,Shi2016}, 
with small modifications to treat Hamiltonians with broken $S^2$ symmetry. Both the energy and
the one-body density matrix (required for
the DMET self-consistency) are computed by the pure estimator, Eq.~(\ref{eq:AFQMCobs}).
We converge the standard deviation of all elements in the one-body density matrix to be less
than 0.001, to make the AFQMC statistical errors (and thus DMET statistical convergence errors) orders of magnitude smaller than
the finite cluster size error. This results in considerably higher statistical accuracy for extensive quantities than typically obtained in the AFQMC literature.

\section{Results} \label{sec:results}
We now present our CDMET and DCA-DMET calculations on the half-filled 1D and 2D Hubbard models,
focusing on the finite-size convergence of the energy and local observables.
As discussed in section~\ref{sec:methods} the DMET correlation potential
preserves $S_z$ symmetry but is allowed to break $S^2$ symmetry. For
the Hubbard models studied here, all the converged self-consistent DMET solutions explicitly break $S^2$ symmetry.
In 1D, we compare our results against exact results from the Bethe Ansatz (BA), while in 2D, we compare to literature benchmark data from AFQMC calculations scaled to the TDL~\cite{Wang2014,Sorella2015,Qin2016}, DMRG calculations scaled to the TDL~\cite{LeBlanc2015}, and iPEPS calculations
scaled to zero truncation error~\cite{Corboz2016}.

\subsection{1D Hubbard model}

We study impurity clusters with $N_\text{imp} = L \leq 24$ sites on a DMET auxiliary lattice with $N=480$ (even $N_c$) or
$N=480 + L$ (odd $N_c$) sites. The auxiliary lattice uses PBC, and as the DCA-DMET impurity Hamiltonian becomes complex for even $N_c$,
we only use auxiliary lattices with an odd $N_c$ in the DCA-DMET calculations. 
We study two couplings $U/t=4$ (moderate coupling) and $U/t=8$ (strong coupling).

\begin{figure}[thpb]
  \centering
  \includegraphics[width=\columnwidth]{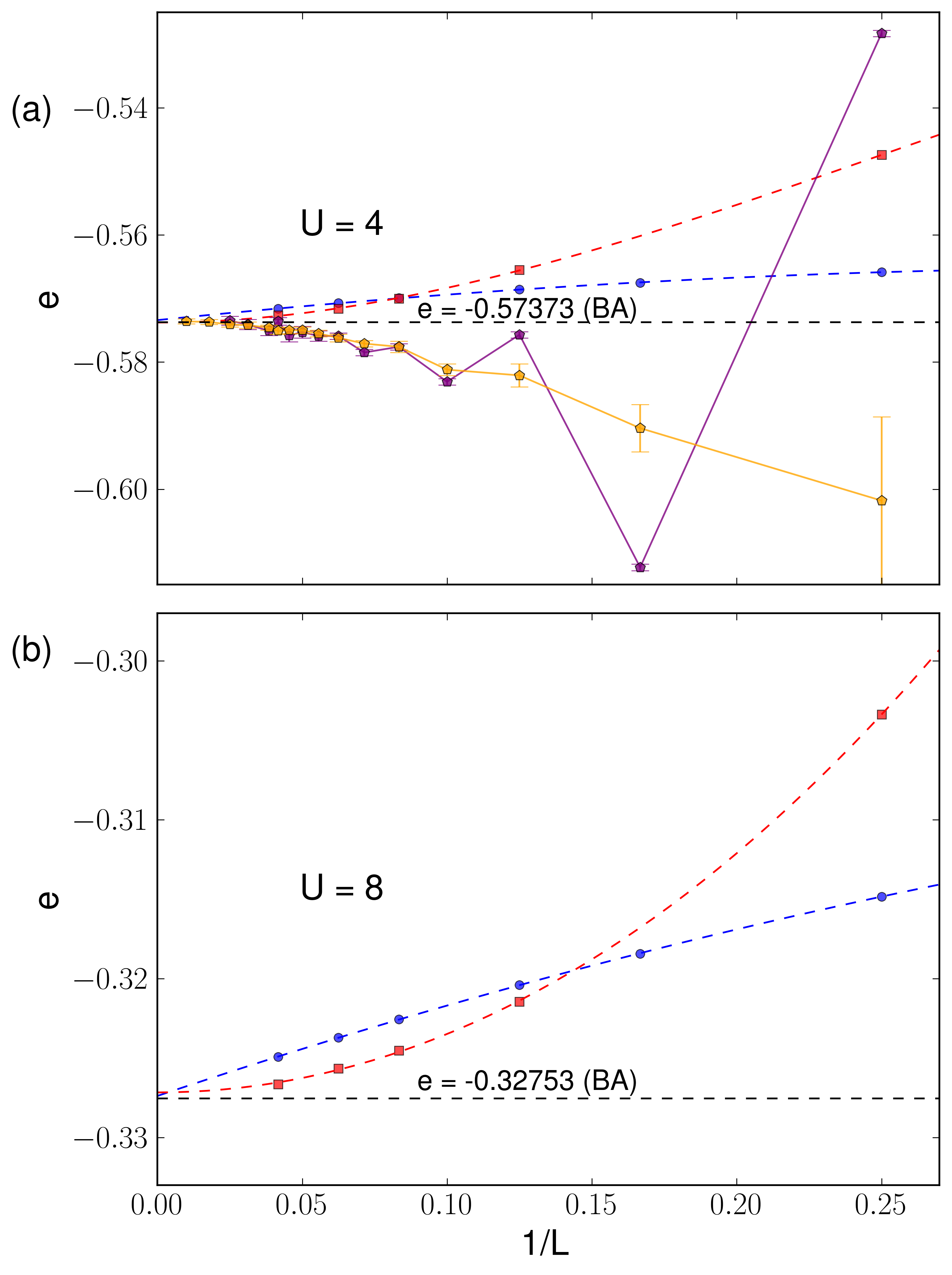}
  \caption{Energy per site, $e$, for the half-filled 1D Hubbard model versus inverse impurity size, $1/L$, 
    from CDMET (blue) and DCA-DMET (red). For comparison, we also plot the same numbers from AFQMC with PBC
    (purple) and TABC (orange) for U/t=4.
    The extrapolations use $e=a+bL^{-1}+cL^{-2}$ for CDMET and $e=a+bL^{-2}+cL^{-3}$ for DCA-DMET.
  (a) U/t=4, (b) U/t=8.}
  \label{fig:energy1D}
\end{figure}

Fig.~\ref{fig:energy1D} shows the energy per site as a function of inverse impurity size $1/L$.  Statistical error bars associated
with the AFQMC solver are not shown
here as they are too small to be visible; this is true for all the CDMET and DCA-DMET results presented in this work.
We extrapolate our finite cluster energy data using the forms presented in Sec.~\ref{sec:finite}. 
As shown in Table~\ref{tab:energy1D}, the extrapolated energies are in generally good agreement with
the exact Bethe ansatz TDL data,
with a deviation of less than $0.001t$. 
To further improve the accuracy, we include the subleading terms in the energy extrapolation, i.e.  $a + b/L + c/L^2$  for CDMET and $a + b/L^2 + c/L^3$
for DCA-DMET (dashed lines in Fig.~\ref{fig:energy1D}). This improves the extrapolated TDL 
results significantly, with the single exception of DCA-DMET at $U/t=8$, where the coefficient of the cubic term is not
statistically significant ($c=0.08(9)$) and the deviation is  already very small.
The subleading terms are more important at $U/t=4$ than at $U/t=8$. This is consistent
with the smaller gap at weaker coupling, that introduces stronger finite size effects.

\begin{table}[thpb]
\centering
\caption{CDMET and DCA-DMET cluster size extrapolation of the energy per site (in units of $t$) for the 1D half-filled Hubbard model.}
\label{tab:energy1D}
\begin{tabular}{|c|l|c|c|}
\hline
\multicolumn{2}{|c|}{extrapolation}                   & U/t=4         & U/t=8 \\      \hline
\multirow{2}{*}{CDMET}    & $a + b/L$                    & -0.5724(3)  & -0.3267(2)  \\ \cline{2-4} 
                          & $a + b/L + c/L^2$       & -0.5734(1)  & -0.3274(1)  \\ \hline
\multirow{2}{*}{DCA-DMET} & $a + b / L^2$                 & -0.5729(4)  & -0.3273(1)  \\ \cline{2-4} 
                          & $a + b/L^2 + c/L^3$        & -0.5738(1)  & -0.3272(1)  \\ \hline
\multicolumn{2}{|c|}{Bethe Ansatz}                    & -0.57373    & -0.32753    \\ \hline
\end{tabular}
\end{table}

To further numerically test the scaling form for the DCA-DMET extrapolation, we include a linear $1/L$ term in the DCA-DMET scaling form,
i.e. $a + b/L + c/L^2$.
While the coefficient of the linear term is statistically significant at $U/t=4$, the extrapolated TDL energy acquires a
larger uncertainty (-0.5749(6)), while for $U/t=8$, the $1/L$ coefficient becomes statistically insignificant ($b=0.003(5)$).
This supports the leading finite-size scaling of the DCA-DMET energy per site as being $O(1/L^2)$.
The finite size scaling of the energy observed for CDMET and DCA-DMET is consistent with similar data
observed for CDMFT and DCA~\cite{Maier2002,Maier2005a}.

In Fig.~\ref{fig:energy1D}(a), we plot the AFQMC results with periodic (PBC) and twist-average (TABC) boundary conditions as well. While PBC energy
oscillate strongly for all cluster sizes, the convergence of TABC is much smoother. The finite-size scaling of bare cluster AFQMC (PBC and TABC) seems quadratic,
which is consistent with the spin-wave theory predictions in 1D~\cite{Huse1988}, and coincides with the scaling of DCA-DMET. Therefore, with large clusters,
the finite-size errors of bare cluster AFQMC and DCA-DMET are comparable and smaller than those of CDMET, while CDMET is much more accurate for small clusters.

%

\begin{figure*}[htpb]
  \subfigure{
    \includegraphics[width=\columnwidth]{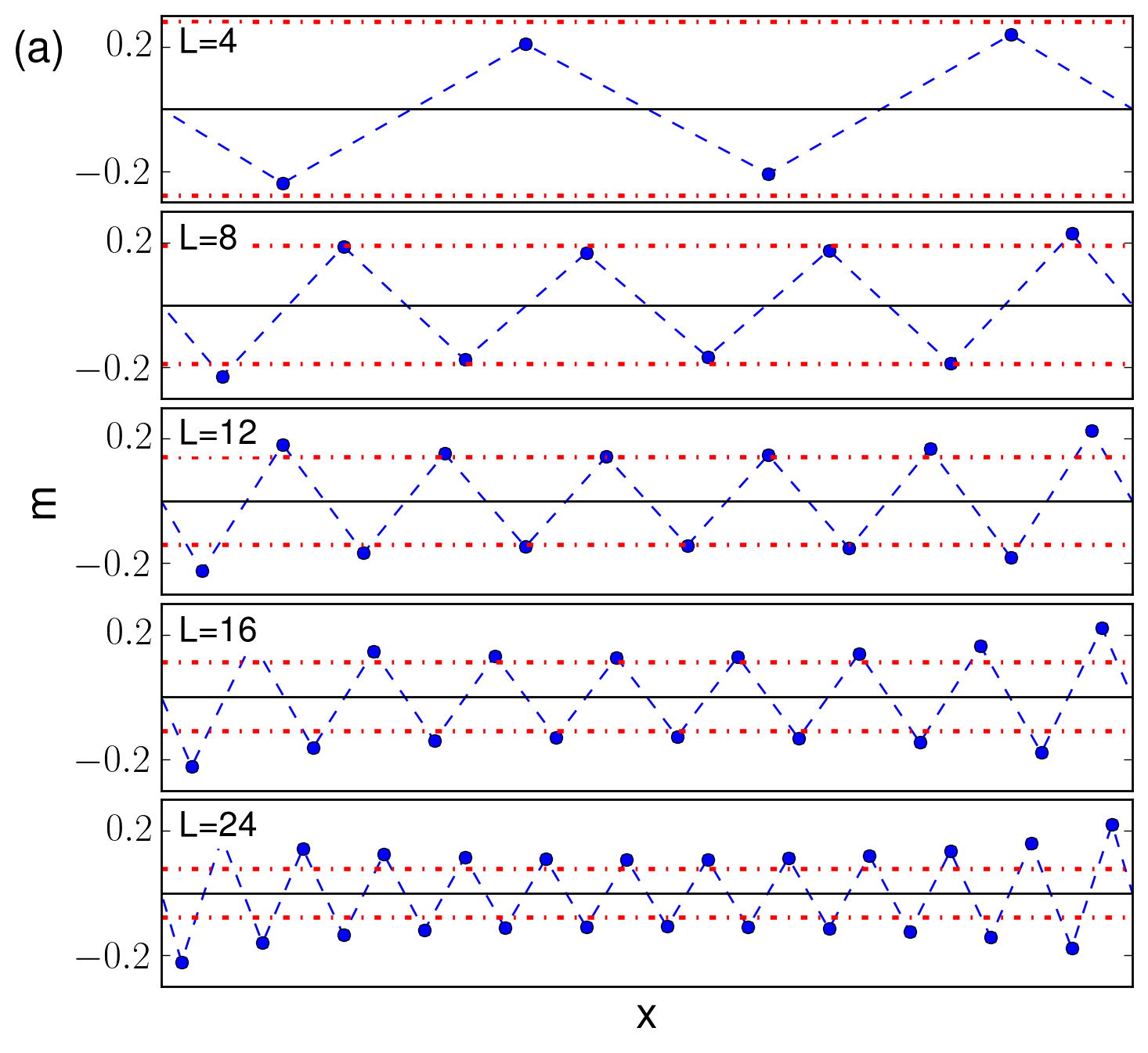}
  }
  \subfigure{
    \includegraphics[width=\columnwidth]{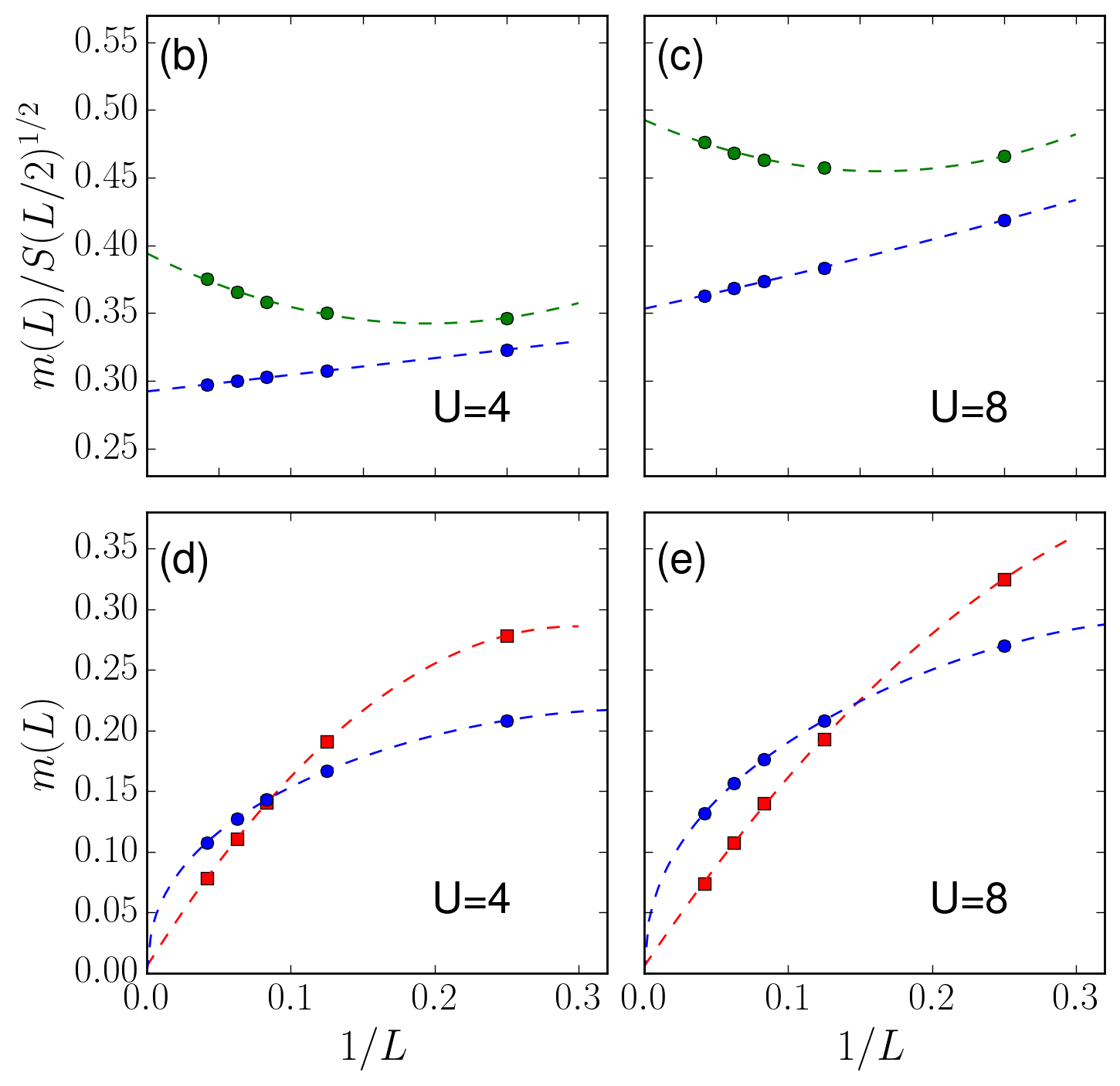}
  }

  \caption{Spin order in the 1D Hubbard model.
  (a) Local spin moments $m$ from  CDMET (blue) and DCA-DMET (red) in finite impurity cluster calculations at $U/t$=4.
  $x$ is the site index scaled to the interval $[0, 1]$ for the CDMET results.
  (b-c) CDMET AF order parameters $m(L)$ divided by spin correlation function $S(L/2)^{1/2}$, versus inverse impurity cluster size $1/L$
  for $U/t=4$ and $U/t=8$ (blue: center average, green: entire cluster average). The extrapolation uses the form $m(L)/S(L/2)^{1/2}=a+bL^{-1}+cL^{-2}$,
  see Eq.~(\ref{eq:1dmscaling}) for details.
  (d-e) DCA-DMET and CDMET (center average) AF order parameters $m(L)$ versus inverse impurity cluster size $1/L$ for $U/t=4$ and $U/t=8$.
  The extrapolation for DCA-DMET values uses the form $m(L)=a+bL^{-1}+cL^{-2}$, see Eq.~(\ref{eq:dcamscaling}) for details.}
  \label{fig:spin1d}
\end{figure*}

We now turn to the spin orders. Although there is no true long-range AF order in 1D, the finite impurity cluster calculations yield
non-zero spin moments, which should extrapolate to zero in the TDL. The local spin moments $m$
are plotted in Fig.~\ref{fig:spin1d}(a).
We see that the spin moments in the CDMET impurity
are largest at the boundary with the AF environment, and decay towards the center. We can understand this
because 
quantum fluctuations
are incompletely treated in the bath orbitals, and thus they are overmagnetized. This effect is propagated to the boundary of the CDMET impurity cluster.
Note that the impurity sites in a DCA-DMET cluster are all
equivalent, and are equally coupled to the environment, resulting in an equal spin magnitude for all sites, to within the statistical error of
the solver. In Fig.~\ref{fig:spin1d}(a) we use the two horizontal lines to represent the spin magnitudes from the DCA-DMET calculations.

To determine the magnetic order parameter, we consider two possible definitions:
(a) the average $|m|$ for the central pair (or the plaquette in 2D); (b) the average $|m|$ over the entire
impurity cluster. These definitions are equivalent for DCA-DMET. In CDMET, they agree in the limit of small clusters ($L=2$) and large clusters ($L\rightarrow\infty$), but differ in between.

The AF order parameters for different cluster sizes are plotted in Figs.~\ref{fig:spin1d}(b)-(e) for different $U$.
The axes uses a logarithmic scale.
For CDMET, we fit the order parameter to the scaling form in Eq.~(\ref{eq:1dmscaling}), up to second order.
The fits are shown in Figs.~\ref{fig:spin1d}(b), (c), and are quite good for both types of measurements. For the average $|m|$ of the central pair,
an almost straight line is observed at both couplings, with the quadratic term close to vanishing ($c=0.00(4)$ for $U/t=4$ and $c=0.12(7)$ for $U/t=8$).
The average $|m|$ over the entire cluster requires a larger $c$ for a good fit. This is because $|m|$ is measured at different points which corresponds
to averaging over different effective lengths $L$ in Eq.~(\ref{eq:1dmscaling}). Averaging over
Eq.~(\ref{eq:1dmscaling}) yields the same leading scaling but introduces more subleading terms.
Overall, the error decreases much more rapidly by using the center average, consistent with  observations in CDMFT~\cite{Biroli2005}.

For DCA-DMET, the scaling form Eq.~(\ref{eq:dcamscaling}) truncated
at second order works well.
This correctly predicts the vanishing local moments at the TDL
($a=0.005(1)$ at $U/t=4$ and $a=0.005(4)$ at $U/t=8$). The $O(1/L)$ scaling of DCA-DMET thus converges faster than
CDMET, whose leading term is $\left(\frac{\sqrt{\log(L/2)}}{L/2}\right)^{1/2}\sim L^{-1/2}$.
While the smallest clusters in CDMET  report a smaller magnetization than seen in DCA-DMET (and 
 thus can be regarded as ``closer'' to the TDL) the cross-over between the DCA-DMET and CDMET moments occurs at smaller clusters
than for the energy itself.

\subsection{2D Hubbard model}

We now show results from the half-filled 2D Hubbard model at $U/t=2,4,6$. We use square impurity clusters of
size $N_{\textrm{imp}}=L\times L$, where for CDMET $L=2,4,6,8,10$ and for
DCA-DMET $L=4,6,8,10$. The $2\times 2$ plaquette is not used in the finite-size scaling  of DCA-DMET as it
is  known from DCA studies to exhibit anomalous behaviour~\cite{Maier2002}, which we also observe.
Also at $U/t=6$, we do not present results for $L=10$, as we are unable to converge the statistical error to high accuracy in the AFQMC
 calculations (within our computational time limits).
The total lattices we used have linear lengths of around $L=120$ ($N=L\times L$), adjusted to fit integer $N_c$, as in the 1D case.

\begin{figure}[thpb]
  \centering
  \includegraphics[width=\columnwidth]{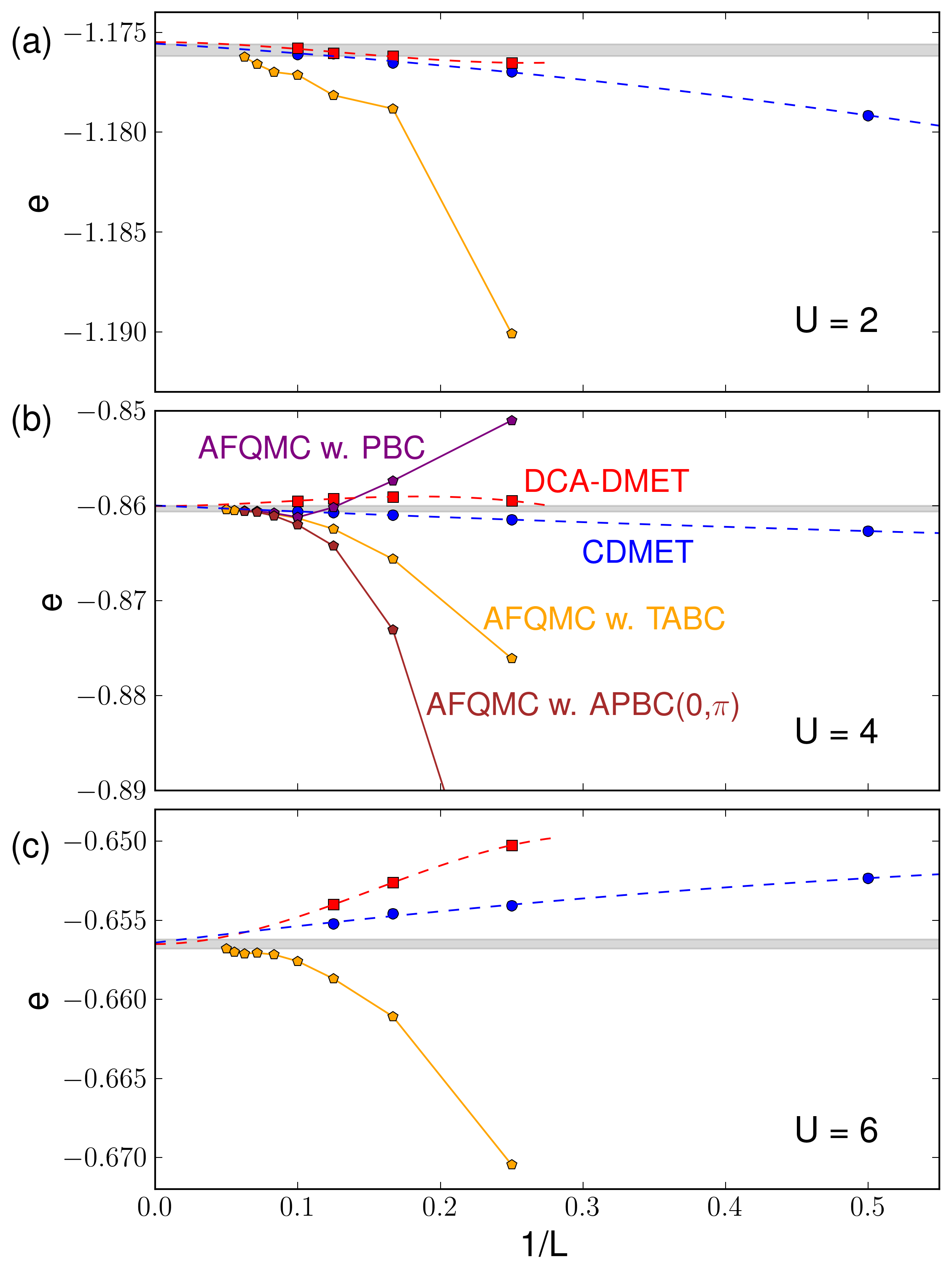}
  \caption{Energy per site $e$ versus $1/L$ in the 2D Hubbard model from CDMET (blue),  DCA-DMET (red) and finite system AFQMC
    (orange: TABC, purple: PBC, brown: APBC for y-direction and PBC for x-direction) (from Ref.~\cite{Qin2016}).
    The consensus range illustrated by the grey-shaded region represents the TDL results of AFQMC, DMRG and iPEPS calculations in 
    Refs.~\cite{LeBlanc2015,Sorella2015,Corboz2016,Qin2016}.
    (a) $U/t=2$. (b)$U/t=4$. (c) $U/t=6$.
  }
  \label{fig:Energy_2D}
\end{figure}

In Fig.~\ref{fig:Energy_2D}, we show the cluster size dependence of the energy per site; the data is tabulated in Table~\ref{tab:Energy_2D}.
Because there are no exact TDL results for the 2D Hubbard model, we show gray ribbons as ``consensus ranges'', obtained from the TDL estimates of
several methods including (i) AFQMC extrapolated to infinite size~\cite{Sorella2015,Qin2016}, (ii) DMRG extrapolated to infinite size~\cite{LeBlanc2015}, and (iii) iPEPS
extrapolated to zero truncation error~\cite{Corboz2016}.
To show the effects of embedding versus bare cluster AFQMC calculations we also plot the AFQMC results of Ref.~\cite{Qin2016}
on finite lattices with up to 400 sites, using TABC for $U/t=2, 4, 6$, as well as 
periodic (PBC) and anti-periodic (APBC) boundary conditions for $U/t=4$.

In 2D, both CDMET and DCA-DMET appear to display much higher accuracy for small clusters, compared to in 1D. Although
DMET is not exact in the infinite dimensional limit, this is similar to the behaviour of DMFT, which improves
with increasing coordination number~\cite{Georges1996}.
The DMET energies for each cluster size are, as expected, much closer to the TDL estimates than the finite system AFQMC energies, even when
twist averaging is employed to reduce finite size effects. For example, the $2\times 2$ CDMET energy is competitive with the $8 \times 8$ AFQMC
cluster energy with twist averaging. This corresponds to several orders of magnitude savings in computation time.
Further, the convergence behaviour generally appears smoother in DMET than with the bare clusters,
likely due to  smaller shell filling effects. This illustrates the benefits of using bath orbitals to
approximately represent the environment in an embedding.

\begin{table*}[thpb]
\centering
\caption{Finite size extrapolation of the energy for the 2D half-filled Hubbard model. }
\label{tab:Energy_2D}
\begin{tabular}{|c|c|c|c|c|c|c|c|c|c|}
\hline
\multirow{2}{*}{methods} &  \multicolumn{2}{|c|}{CDMET} &  \multicolumn{2}{|c|}{DCA-DMET} & \multicolumn{2}{|c|}{AFQMC} & \multirow{2}{*}{DMRG~\cite{LeBlanc2015}} 
& \multirow{2}{*}{iPEPS~\cite{Corboz2016}} & \multirow{2}{*}{Consensus range} \\\cline{2-7}
      &  $a + b/L$  &  $a + b/L + c/L^2$ &  $a + b/L^2$ &  $a + b/L^2 + c/L^3$  & TABC~\cite{Qin2016} & MBC~\cite{Sorella2015} &            &           &            \\ \hline
U/t=2 &  -1.1752(1) &  -1.1756(3)        &  -1.1758(1)  &  -1.1755(2)           & -1.1760(2)          & -1.17569(5)            & -1.176(1)  &  -        & -1.1758(3) \\ \hline
U/t=4 &  -0.8601(1) &  -0.8600(1)        &  -0.8593(2)  &  -0.8600(2)           & -0.8603(2)          & -0.86037(6)            & -0.8605(5) & -0.8603(5)& -0.8603(3) \\ \hline
U/t=6 &  -0.6560(2) &  -0.6564(6)        &  -0.6550(4)  &-0.6565\footnotemark[1]& −0.6567(3)          & -                      & -0.6565(1) &  -        & -0.6565(3) \\ \hline
\end{tabular}
\footnotetext[1]{uncertainty cannot be computed due to insufficient data points in the fit.}
\end{table*}

\begin{table*}
  \centering
  \caption{Estimated staggered magnetization for the 2D half-filled Hubbard model at TDL.}
  \label{tab:AF_2D}
  \begin{tabular}{|c|c|c|c|c|c|c|}
    \hline
    methods & CDMET    & DCA-DMET & DQMC~\cite{Varney2009} & Pinning field QMC~\cite{Wang2014} & AFQMC w. TABC~\cite{Qin2016} & AFQMC w. MBC~\cite{Sorella2015}\\\hline
    U/t=2   & 0.115(2) & 0.120(2) & 0.096(4)               & 0.089(2)                          & 0.119(4)                     & 0.120(5)                       \\\hline
    U/t=4   & 0.226(3) & 0.227(2) & 0.240(3)               & 0.215(10)                         & 0.236(1)                     & -                              \\\hline
    U/t=6   & 0.275(8) & 0.261\footnotemark[1]& 0.283(5)   & 0.273(5)                          & 0.280(5)                     & -                              \\\hline
  \end{tabular}
  \footnotetext[1]{uncertainty cannot be computed due to insufficient data points in the fit.}
\end{table*}

We extrapolate the DMET finite cluster results to obtain  TDL estimates. As in the 1D Hubbard model, we use the scaling forms proposed
in section~\ref{sec:finite}, i.e. $a + b/L (+ c/L^2)$  for CDMET and
 $a + b/L^2 (+ c/L^3)$ for DCA-DMET.
The results are summarized in Table~\ref{tab:Energy_2D} and plotted in Fig.~\ref{fig:Energy_2D}.
The TDL energy estimates fall within the TDL consensus range, with an error bar competitive with the best large-scale ground state calculations.
The DMET estimates are also all in agreement (within 2$\sigma$) of our earlier CDMET extrapolations that only used clusters of up to $4\times 4$ sites in Refs.~\cite{Zheng2016,LeBlanc2015}. The largest deviation from our  earlier small cluster DMET extrapolations is for $U/t=2$ where finite size effects are strongest; the current estimates of $-1.1756(3)$ (CDMET) and $-1.1755(2)$ (DCA-DMET)
can be compared with our small cluster estimate of $-1.1764(3)$, and
 the recent TDL estimate of Sorella of $-1.17569(5)$, obtained by extrapolating AFQMC energies from clusters as large as 1058 sites, using modified boundary conditions~\cite{Sorella2015}.
Note that the subleading terms are more important for accurate extrapolations in 2D than they are in 1D.
This is simply because
we do not reach as large linear dimensions in 2D as in 1D, which means that we are not fully in the asymptotic regime.
For the same reason it is more difficult to see the crossover between the convergence of DCA-DMET and CDMET. For $U/t=2$, it appears advantageous to use the DCA-DMET formulation
already for clusters of size $L\ge 4$, while at $U/t=4, 6$ it appears necessary to go to clusters larger than the largest linear size used
in this study, $L=10$.

\begin{figure}[thpb]
  \centering
  \includegraphics[width=\columnwidth]{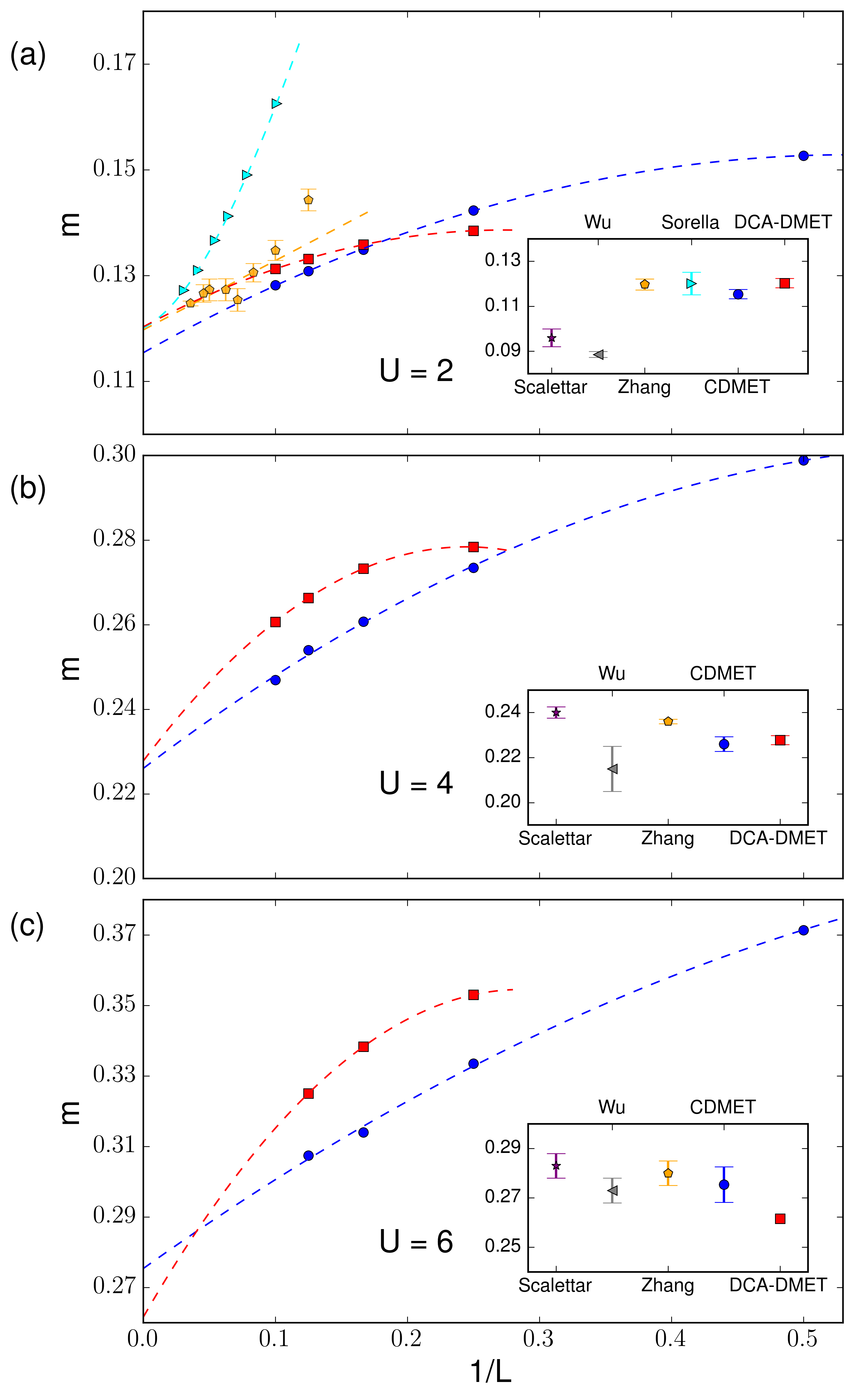}
  \caption{Antiferromagnetic order parameter $m$ versus $1/L$ in the 2D Hubbard model from CDMET (blue), DCA-DMET (red) and finite system AFQMC using TABC~\cite{Qin2016} (orange)
    and modified boundary conditions~\cite{Sorella2015} (cyan). The DMET results extrapolate to the TDL uses the form $m(L) = a + bL^{-1} + cL^{-2}$.
    Insets: CDMET and DCA-DMET TDL estimates with errorbars including fitting and AFQMC statistical
    uncertainties, compared to the determinantal Monte Carlo simulations by Scalettar and coworkers~\cite{Varney2009}, 
    pinning field QMC simulations by Wu and coworkers~\cite{Wang2014},
    AFQMC with TABC by Qin et. al.~\cite{Qin2016} and the modified boundary conditions by Sorella~\cite{Sorella2015}.
  (a) $U/t=2$. (b) $U/t=4$. (c) $U/t=6$.}
  \label{fig:AForder}
\end{figure}

The AF order in the half-filled 2D Hubbard model is long-ranged in the ground state.
In Fig.~\ref{fig:AForder}, the AF order parameters from DMET are plotted and extrapolated, with insets showing comparisons of TDL estimates
with the other methods. In addition, we summarize the extrapolated TDL estimates for the AF order parameters in Table.~\ref{tab:AF_2D}.
For CDMET, the order parameters are measured as the average magnitude of 
the central plaquette. We fit the magnetization data to the form suggested in Section~\ref{sec:finite}, i.e.
$a+ b/L + c/L^2$ for both CDMET and DCA-DMET.
These fits lead to good agreement between the CDMET and DCA-DMET TDL estimates, supporting the scaling form used.
At $U/t=4$, the CDMET and DCA-DMET TDL  moments are in good agreement with the estimates from two different AFQMC calculations,
with competitive error bars. At $U/t=6$, the CDMET TDL moment is consistent with the two AFQMC estimates
and the DCA-DMET estimate, although the DCA-DMET estimate is
somewhat smaller than the two AFQMC estimates. (We do not have
errors bars for the $U/t=6$ DCA-DMET moment as we are fitting 3 data points to a 3 parameter fit).

The TDL magnetic moment at $U/t=2$ is an example for which current literature estimates are in disagreement. While earlier AFQMC calculations in 
Ref.~\cite{Varney2009,Wang2014,LeBlanc2015}
appear to give an estimate close to $m\sim 0.09$, the AFQMC estimates from recent work of Sorella~\cite{Sorella2015} and Qin et al~\cite{Qin2016,afqmc_note}
using larger clusters and modified and twist average boundary conditions
predict a moment of
$m \sim 0.120(5)$ and $0.119(4)$, respectively. This is much closer
to our earlier DMET result of $m \sim 0.133(5)$ extrapolated
from small clusters of up to $4\times 4$ in size.
Revising this with the larger CDMET and DCA-DMET clusters
in this work we can now confirm the larger value of the TDL magnetic moment,
$m \sim 0.115(2)$ (CDMET) and $m \sim 0.120(2)$ (DCA-DMET) with very small error bars.
The underestimate of the moment seen in earlier QMC work is likely due to the non-monotonic
convergence of the moment with cluster size when using PBC, as identified in Sorella's work~\cite{Sorella2015}.
In contrast to PBC calculations and the TABC calculations shown here (orange) which display some scatter,
the dependence on cluster size is very mild once embedding is introduced.
This once again highlights the ability of the embedded approach to capture
some of the relevant aspects
even of long-wavelength physics, leading to good convergence of local observables.

\section{Conclusions}

In this work, we carried out a detailed study of the cluster size convergence of density matrix embedding theory,
using an auxiliary field quantum Monte Carlo solver (AFQMC) in order to reach larger cluster sizes than
studied before. In addition to
the original cluster density matrix embedding formulation (CDMET), we introduced a ``dynamical cluster'' variant (DCA-DMET)
that restores translational invariance in the impurity cluster and accelerates finite size convergence. 
Using the half-filled one- and two-dimensional Hubbard models where AFQMC has no sign problem, as examples,
we numerically explored the finite size convergence of the energy and the magnetization.
The energy convergence of CDMET and DCA-DMET goes like $O(1/L)$ and $O(1/L^2)$ respectively, where $L$ is the linear dimension
of the cluster, similar to that observed in cellular dynamical mean-field theory and the dynamical cluster approximation.
The convergence of the magnetization follows a scaling relation related to the magnetic correlation function,
with the DCA-DMET converging more quickly than CDMET.
In the case of the 2D Hubbard model, our thermodynamic limit extrapolations from
both CDMET and DCA-DMET are competitive with the most accurate estimates
in the literature, and in the case of $U/t=2$ where finite size effects are particularly strong,  help to
determine the previously uncertain magnetic moment.

In all the cases we studied here, the use of density matrix embedding, as compared to computations using bare clusters with any form of boundary condition,
decreased the computational cost required to obtain
a given error from the TDL significantly, sometimes by orders of magnitudes.
Since the computational scaling of the AFQMC solver employed here is quite modest with cluster size (cubic)
this improvement would only be larger when using other, more expensive solvers.

The availability of a DCA formulation now presents two options for how to perform cluster DMET calculations.
The DCA-DMET formulation appears superior for large clusters due to the faster asymptotic convergence,
however, it is typically less accurate for small clusters than CDMET.
When performed in conjunction, the consistency of TDL estimates from CDMET and DCA-DMET serves as a strong check on the reliability of the
DMET TDL extrapolations.

\begin{acknowledgments}
We thank Mingpu Qin for helpful communications and assistance.
This work was supported by the US Department of Energy, Office of Science (Bo-Xiao Zheng by Grant No. DE-SC0010530; 
Joshua Kretchmer and Garnet Kin-Lic Chan by Grant No. DE-SC0008624; Hao Shi and Shiwei Zhang by Grant No. DE-SC0008627) and by the Simons Foundation. 

%
\end{acknowledgments}

\appendix
\section{Constraints for sign-problem free correlation potentials in DMET} \label{sec:phsymm}
We first motivate our derivation by recalling how AFQMC becomes sign-problem free in the half-filled Hubbard model
on a bipartite lattice. Given the repulsive Hubbard model with chemical potential $\mu=U/2$
\begin{equation}
  H-\mu n=-t\sum_{\langle ij\rangle\sigma} a_{i\sigma}^{\dagger}a_{j\sigma} + U\sum_{i}[n_{i\uparrow}n_{i\downarrow}-\frac{1}{2}(n_{i\uparrow}+n_{i\downarrow})]
\end{equation}
we perform the partial particle-hole transformation on \textit{only} the spin-up electrons
\begin{equation}
  \hat{P}: a_{i\uparrow}^{\dagger}\rightarrow (-)^ia_{i\uparrow}, a_{i\uparrow}\rightarrow (-)^ia_{i\uparrow}^{\dagger}
  \label{eq:ph_transform}
\end{equation}
where the parity term $(-)^i$ is $1$ for sublattice $A$, and $-1$ for the other sublattice, $B$. The transformation results in the attractive Hubbard model
\begin{equation}
  \hat{P}H\hat{P}^{-1}=-t\sum_{\langle ij\rangle,\sigma}a_{i\sigma}^{\dagger}a_{j\sigma} - U\sum_{i}[n_{i\uparrow}n_{i\downarrow}-\frac{1}{2}(n_{i\uparrow}+n_{i\downarrow}-1)]
\end{equation}
which is well-known to be sign-problem free at any occupation. This is seen by performing the Hubbard-Stratonovich transformation,
where the Trotter propagator becomes~\cite{blankenbecler1981monte}, 
\begin{equation}
  e^{- \tau\hat{P}H\hat{P}^{-1}} \approx
  \exp(\tau t\sum_{ij\sigma} a_{i\sigma}^{\dagger}a_{j\sigma})\prod_i\sum_{x_i=\pm1}\frac{1}{2}e^{\gamma x_i(n_{i\uparrow}+n_{i\downarrow}-1)}
  \label{eq:projector_Hubbard}
\end{equation}
with $\gamma=\cosh^{-1}e^{\tau U/2}$. Notice that  Eq.~(\ref{eq:projector_Hubbard}) is spin-symmetric, thus as long as the trial wavefunction $|\Phi_t\rangle$
is  spin-symmetric, the walkers $|\Phi_w\rangle$ are also spin-symmetric. The overlap
\begin{equation}
  \langle \Phi_t|\Phi_w\rangle = \langle \Phi_{t\uparrow}|\Phi_{w\uparrow}\rangle \langle \Phi_{t\downarrow}|\Phi_{w\downarrow}\rangle 
  = |\langle\Phi_{t\uparrow}|\Phi_{w\uparrow}\rangle|^2 \ge 0
\end{equation}
then eliminates the sign problem. From this argument, we also see why the repulsive Hubbard model is sign problem free only at half-filling,
since we require the same number of spin-up holes and spin-down particles in the wavefunction.

In DMET calculations, it is easy to show that if the partial particle-hole symmetry is preserved in the lattice Hamiltonian, the resulting impurity
problem remains sign-problem free.
Consider the partial particle-hole transformation,
Eq.~(\ref{eq:ph_transform}), acting on the non-interacting lattice Hamiltonian in Eq.~(\ref{eq:h_mf}), with chemical potential $\mu=U/2$
\begin{equation}
  \begin{split}
  &\hat{P}(h-\mu n)\hat{P}^{-1} \\
  &=\hat{P}[h_0+u-\sum_i\frac{U}{2}(n_{i\uparrow}+n_{i\downarrow})]\hat{P}^{-1}\\
  &=h_0+N_c(\sum_{i\in C}u_{ii\uparrow}-UN_{\text{imp}}/2)+\\
  &\sum_C\sum_{i,j\in C}\{[\frac{U}{2}\delta_{ij}-(-)^{i+j}u_{ij\uparrow}]a_{i\uparrow}^{\dagger}a_{j\uparrow}+(u_{ij\downarrow}
  -\frac{U}{2}\delta_{ij})a_{i\downarrow}^{\dagger}a_{j\downarrow}\}
  \end{split}
  \label{eq:h_mf_transformed}
\end{equation}

To impose spin symmetry, we have
\begin{equation}
  \frac{U}{2}\delta_{ij}-(-)^{i+j}u_{ij\uparrow} = u_{ij\downarrow}-\frac{U}{2}\delta_{ij}
\end{equation}
which leads to Eq.~(\ref{eq:phsymm}). When this condition is satisfied, the ground state of the transformed lattice
Hamiltonian $\hat{P}(h-\mu n)\hat{P}^{-1}$ 
is a spin-symmetric Slater determinant and thus the bath orbitals obey $R^\uparrow=R^\downarrow$. The impurity model Hamiltonian $h_{\text{imp}}$ (Eq.~(\ref{eq:h_imp})) is thus
sign-problem free, as $\bar{h}$ is clearly spin-symmetric and $V_{\text{imp}}$ transforms to an attractive Hubbard interaction.

Note that our argument applies to both CDMET and DCA-DMET, since the DCA transformation preserves the partial particle-hole symmetry, which is
the only structure assumed of $h_0$ in the above derivation.

\section{Symmetries in the DCA-DMET correlation potential}\label{sec:trans_symm}

We here consider translational symmetry in the correlation potential in the presence of antiferromagnetic order.
Instead of the normal translational operators, the lattice
Hamiltonian is invariant under the spin-coupled translational operators
\begin{equation}
  T_x: a_{i\sigma}^{(\dagger)} \rightarrow 
  \begin{cases}
    a_{i+x,\sigma}^{(\dagger)},& \text{if } x \text{ is even}\\
    a_{i+x,\bar\sigma}^{(\dagger)},& \text{if } x \text{ is odd}
  \end{cases}
\end{equation}
where the parity of $x$ represents whether a translation brings a site to the same or different sublattice. The Hubbard Hamiltonian is
invariant under $T_x$ operations, because it has both translational and time-reversal symmetry.
Transforming the correlation potential with the spin-coupled translational operators yields
\begin{widetext}
  \begin{equation}
  \begin{split}
  &\text{for even } x,
  T_x u T_x^{-1}
  =\sum_C\sum_{i,j\in C}\sum_\sigma u_{ij\sigma}a_{i+x\sigma}^{\dagger}a_{j+x\sigma}
  =\sum_C\sum_{i,j\in C}\sum_\sigma u_{i-x,j-x,\sigma}a_{i\sigma}^{\dagger}a_{j\sigma} = u\\
  &\text{for odd } x,
  T_x u T_x^{-1}
  =\sum_C\sum_{i,j\in C}\sum_\sigma u_{ij\sigma}a_{i+x\bar\sigma}^{\dagger}a_{j+x\bar\sigma}
  =\sum_C\sum_{i,j\in C}\sum_\sigma u_{i-x,j-x,\bar\sigma}a_{i\sigma}^{\dagger}a_{j\sigma} = u
  \end{split}
\end{equation}
\end{widetext}
leading to the constraint
\begin{equation}
  u_{ij\sigma}=
  \begin{cases}
    u_{0,j-i,\sigma}, &\text{if } i \text{ is even}\\
    u_{0,j-i,\bar\sigma}, &\text{if } i \text{ is odd}
  \end{cases}
  .
\end{equation}
This constraint, as one can easily verify, is compatible with the partial particle-hole symmetry required for sign-free AFQMC simulations
in the Hubbard model.  
\bibliographystyle{apsrev4-1}
\bibliography{ref}

\begin{thebibliography}{43}
\bibitem [{afq()}]{afqmc_note}%
  \BibitemOpen
  \href@noop {} {}\bibinfo {note} {The AFQMC result of antiferromagnetic order
  parameter at U/t=2 in Ref. 14 has an error in the extrapolation to the TDL,
  which was corrected in Ref. 18}\BibitemShut {NoStop}%
\end{thebibliography}
\end{document}